\documentclass[conference]{IEEEtran}
\IEEEoverridecommandlockouts
\usepackage{cite}
\usepackage{fancyhdr}

\usepackage{rotating} 
\usepackage{helvet}
\usepackage{booktabs}   
\usepackage{makecell}
\usepackage{hyperref}
\usepackage{url}
\usepackage{pifont}
\usepackage{listings}
\usepackage{tcolorbox}
\usepackage{graphicx}
\usepackage{utfsym}
\usepackage{amsmath}
\usepackage{subfigure}
\usepackage{hyperref}
\usepackage[linesnumbered,ruled,vlined]{algorithm2e}
\usepackage{multirow}

\newcommand{\wahib}[1] {{\color{blue}{#1}}}

\usepackage{color}
\usepackage{makecell}
\usepackage{tablefootnote}
\usepackage{hhline}
\usepackage{wrapfig}

\usepackage {captdef}
 \usepackage{amssymb}
\usepackage[table,xcdraw]{xcolor}
\usepackage{inconsolata}
\definecolor{dkgreen}{rgb}{0,0.6,0}
\definecolor{gray}{rgb}{0.5,0.5,0.5}
\definecolor{mauve}{rgb}{0.58,0,0.82}

\usepackage[skip=1pt]{caption}

\usepackage{listings}
\usepackage{xcolor}
\usepackage{tikz}

\usepackage{listings}
\usepackage{xcolor}

\usepackage{lipsum}

\usepackage{tikz}
\usetikzlibrary{calc}
\usepackage{pgfplots}
\pgfplotsset{compat=1.18}

\def\BibTeX{{\rm B\kern-.05em{\sc i\kern-.025em b}\kern-.08em
    T\kern-.1667em\lower.7ex\hbox{E}\kern-.125emX}}

\usetikzlibrary{decorations.pathreplacing,angles,quotes}

\newcommand{\method}{\texttt{SparkleDock}}

\usepackage{cite}
\usepackage{hyperref}
\usepackage[switch]{lineno} 

\hypersetup{hidelinks}
\begin{document}

\title{Scalable High-Fidelity Macromolecular Docking for GPU-Accelerated Supercomputers}
\author{
\IEEEauthorblockN{
Xiangyu Meng\textsuperscript{1,2,\textdagger},
Peng Chen\textsuperscript{3,\textdagger},
Mingzhen Li\textsuperscript{4},
Jianmin Wang\textsuperscript{5},
Sen Wang\textsuperscript{1},
Guangming Tan\textsuperscript{4},
Weile Jia\textsuperscript{4},\\
Mohamed Wahib\textsuperscript{3,*},
Tao Luo\textsuperscript{2},
Xun Wang\textsuperscript{1,*}
}

\IEEEauthorblockA{\textsuperscript{1}
College of Computer Science and Technology,
Shandong Key Laboratory of Intelligent Oil \& Gas Industrial Software,\\
China University of Petroleum (East China), Qingdao, China
}

\IEEEauthorblockA{\textsuperscript{2}
A*STAR Institute of Advanced Intelligence and Computing, Singapore
}

\IEEEauthorblockA{\textsuperscript{3}
RIKEN Center for Computational Science, Kobe, Japan
}

\IEEEauthorblockA{\textsuperscript{4}
Institute of Computing Technology, Chinese Academy of Sciences,
Beijing, China
}

\IEEEauthorblockA{\textsuperscript{5}
Department of Computer Science and Engineering,
The Chinese University of Hong Kong, China
}

\IEEEauthorblockA{
x\_meng0420@163.com, peng.chen@riken.jp,
limingzhen@ict.ac.cn, jmwang113@hotmail.com,
homletree@s.upc.edu.cn,\\tgm@ict.ac.cn,
jiaweile@ict.ac.cn, mohamed.attia@riken.jp,
luo\_tao@a-star.edu.sg, wangsyun@upc.edu.cn
}
\thanks{\textsuperscript{\textdagger}Co-first authors; Corresponding authors: Xun Wang and Mohamed Wahib}
}

\maketitle 

\thispagestyle{fancy}
\lhead{}
\rhead{}
\chead{}
\lfoot{\footnotesize{
SC26, November 15-20, 2026, Chicago, Illinois, USA
\newline 979-8-3195-4789-7/26/\$31.00 \copyright 2026 IEEE}}
\rfoot{}
\cfoot{}
\renewcommand{\headrulewidth}{0pt}
\renewcommand{\footrulewidth}{0pt}

\begin{abstract}
Flexible macromolecular docking offers high-fidelity predictions of biomolecular interactions, but remains prohibitively expensive at scale. Among existing approaches, LightDock leverages Glowworm Swarm Optimization (GSO) for accuracy, yet suffers from limited parallelism, irregular computation, and severe load imbalance, preventing efficient execution on GPU supercomputers. We present SparkleDock, a scalable GSO-based docking framework enabling near-real-time flexible docking. 
We redesign GSO to expose massive fine-grained parallelism at the glowworm-agent level, and restructure the dominant energy scoring computation into a Tensor Core-compatible formulation, enabling efficient execution of irregular pairwise interactions through structured matrix operations. We further introduce a performance-model-driven scheduling for load balancing and out-of-core scaling across GPUs. SparkleDock achieves 9.7× and 18.9× speedups over LightDock on single A100 and H100 GPU, and delivers over two orders of magnitude acceleration at scale. On 512 GPUs, it reduces docking time from hours to seconds, enabling large-scale, high-fidelity virtual screening previously impractical with flexible docking.

\begin{IEEEkeywords}
Macromolecular docking, GPU-accelerated supercomputer, Tensor cores, Strong scaling
\end{IEEEkeywords}

\end{abstract}


\section{Introduction}

Flexible macromolecular docking is essential for accurately modeling biomolecular interactions, including proteins~\cite{de2012protein,stelzl2005human,wang2024exploring}, antibody–antigen~\cite{kilambi2017structure,lu2020development}, membrane receptor–soluble protein~\cite{roel2020integrative} interactions, and drug design~\cite{antunes2015understanding,sledz2018protein}. Compared to rigid-body methods, it improves the docking success rate by about 20\%–30\% as it accounts for conformational flexibility that traditional rigid-body docking cannot capture.
Among existing flexible docking algorithms, LightDock adopts a Glowworm Swarm Optimization (GSO) method~\cite{jimenez2018lightdock,roel2020lightdock,roel2020integrative} that leverages multiple intelligent agents (i.e. glowworm agents) to optimize and search for flexible docking poses with stable binding energy. It has gained wide adoption for its high docking accuracy with an efficient and intelligent sampling approach.
While LightDock achieves high docking fidelity in comparison to other docking approaches (as seen in Tab.~\ref{tab:compare}), the algorithmic characteristics of GSO limits its scalability and hinder the efficient utilization of modern GPU architectures and large-scale distributed systems.
Applying such applications for virtual screening on massive protein datasets like UniProt~\cite{uniprot2015uniprot} would take several years.
This obstacle highlights the importance of developing high-performance solutions for large-scale flexible macromolecular docking.


The swarm-based LightDock exhibits three key bottlenecks: limited parallelism, irregular computation patterns, and inherent load imbalance, all of which hinder efficient utilization of GPU architectures and large-scale distributed systems. 

\textbf{Limited parallelism.} LightDock exposes parallelism primarily at the swarm level, where each swarm is processed independently. 
However, GSO inherently organizes the optimization as a set of interacting glowworm agents within each swarm, where each glowworm agent updates its state based on its local neighborhood~\cite{wu2012improvement,krishnanand2009glowworm}.
This swarm-level design fundamentally limits the degree of parallelism to the number of swarms (typically hundreds to thousands), far below the massive concurrency supported by modern GPUs. 
Moreover, the iterative dependency among agents within each optimization step further restricts parallel execution.

\textbf{Irregular compute pattern.} The energy score calculation dominates the computational cost, accounting for 89\% of the total runtime and over 95\% of the total FLOPs. 
This computation involves evaluating full pairwise interactions between atoms for each glowworm agent, resulting in quadratic complexity with respect to the number of atoms. 
Unlike structured dense linear algebra workloads, this irregular pairwise Euclidean distance computation deviates from structured matrix operations and cannot be directly mapped onto hardware designed for dense linear algebra like Tensor Core Units (TCU)~\cite{markidis2018nvidia}, necessitating a careful reformulation.

\textbf{Poor scalability.} 
GSO exhibits inherent load imbalance due to heterogeneous exploration of the conformational space, where different glowworm agents incur unequal computational costs. 
This imbalance is further amplified by swarm-level parallelism, which processes swarms independently without fine-grained workload balancing. 
Consequently, existing implementations exhibit low GPU utilization and fail to scale efficiently across multiple GPUs or distributed systems for high-throughput screening. 
In addition, large-scale docking tasks (e.g. PDB entry 4GAM) introduce substantial memory demands, often reaching hundreds of gigabytes, which exceed the memory capacity of individual GPUs. For instance, docking PDB entry 4GAM generates hundreds of thousands of candidate glowworm agents and corresponding complexes. Each agent evaluates approximately 19.5 million receptor–ligand atom pairs per simulation, collectively resulting in a memory footprint of hundreds of gigabytes.
These factors collectively hinder scalable deployment, necessitating coordinated load balancing and out-of-core execution strategies.

This paper presents \method{}, a scalable GSO macromolecular docking framework, which exploits TCU acceleration at the GPU level, and further scales on GPU-accelerated supercomputers. The design of \method{} is motivated by the growing demand to bridge the gap between the high-accuracy flexible macromolecular docking algorithm and the inherent performance and scaling limitations of GSO.
Experimental results on single A100 and H100 GPU demonstrate that \method{} achieves 9.7$\times$ and 18.9$\times$ speedups, respectively, in comparison to LightDock. The scaling evaluation shows that \method{} achieves effective strong scaling on multi-GPU systems with up to 512 GPUs, approaching effective strong scaling, particularly for larger workloads. By overcoming these scalability barriers, \method{} enables large-scale virtual screening and time-expensive docking workloads that were previously impractical with existing flexible docking methods.
The contributions of this paper are as follows:
{
\setlength{\leftmargini}{15 pt}
    \begin{itemize}
    \item \textit{Novel algorithmic design of GSO to expose parallelism}. We construct a fine-grained, glowworm agent-level parallelization scheme to match the compute hierarchy and memory subsystem of GPUs. 
    \item \textit{Co-design of the energy scoring computation with TCUs}. We restructure the dominant energy scoring computation into a TCU–compatible formulation (i.e. 89.1\% of the total execution time), transforming it into structured matrix computations to fully exploit TCUs while preserving the original scoring semantics. 
    \item \textit{Scalability}. We develop performance model guided scaling methods in \method{} for efficient load balance and chunk division under different computing scales. 
\end{itemize}
}

The rest of this paper is organized as follows.
Section~\ref{sec:background} reviews the background and related work. 
Section~\ref{sec:design} presents the proposed \method{} algorithm.
Section~\ref{sec:evaluation} reports the experimental results. 
Finally, Section~\ref{sec:conclusion} concludes the paper.

\begin{algorithm}[!t]
\small
\caption{Macromolecular docking using GSO}
\label{alg:lightdock}
\SetKwInOut{Input}{Input}
\SetKwInOut{Output}{Output}
\SetKwComment{Comment}{// }{}
\SetKwFunction{PreparePose}{PreparePose}
\SetKwFunction{CalcEnergyScore}{CalcEnergyScore}
\SetKwFunction{CalcNeighbors}{CalcNeighbors}
\SetKwFunction{CalcMovement}{CalcMovement}
\SetKwFor{ForPar}{for}{do in parallel}{endfor}
\SetKwComment{Comments}{ $\triangleright$}{}
\Input{
Receptor backbone $\mathcal{R}$; Ligand backbone $\mathcal{L}$; distance-energy lookup table $\theta$
}

\Output{Glowworm agent vectors $\mathcal{G}$}

$I \gets$ \textsc{InitSwarm}($\mathcal{R}$); \label{alg1:line1}\Comments{\textcolor{blue}{Swarms Initialization}}
$\mathcal{G} \gets$ \textsc{InitGlowworm}($\mathcal{R},\mathcal{L},I$); \label{alg1:line2}\Comments{\textcolor{blue}{$\mathcal{G}$ Initialization}}
{\color{blue}\tcp*[h]{ Simulation} \label{alg1:line3}}\\ 
\ForPar{$i \gets 0$ \KwTo $N_s-1$}{ 
    \For{$t \gets 0$ \KwTo $S-1$}{
        \For{$j \gets 0$ \KwTo $N_g-1$}{ 
            {\color{blue}\tcp*[h]{ Prepare docking Pose}} \\
            \For{$x \gets 0$ \KwTo $N_r-1$}{ \label{alg1:line7}
             $\mathcal{A}_{x}^{(i,j)} \gets$ \textsc{UpdateRecPos} ($\mathcal{R}_{x},\mathcal{G}^{(i,j)}$)\;
            }
            \For{$x \gets 0$ \KwTo $N_l-1$}{ 
             $\mathcal{B}_{x}^{(i,j)} \gets$ \textsc{UpdateLigPos} ($\mathcal{L}_x,\mathcal{G}^{(i,j)}$)\;
            }\label{alg1:line12}
            {\color{blue}\tcp*[h]{ Calc DFIRE score(Bottleneck)}}\\ \label{alg1:12}
            $\mathcal{D} \gets$ \textsc{CalcPairwiseDist} ($\mathcal{A}^{(i,j)}, \mathcal{B}^{(i,j)}$)\; \label{alg1:14}
            $Idx \gets$ \textsc{DistBinning} ($\mathcal{D}$)\; \label{alg1:15}
            $E^{(i,j)} \gets$ \textsc{Accumulation} ($Idx,\theta$)\; \label{alg1:16}
            {\color{blue}\tcp*[h]{ Calc neighbors $l$ \& Movement}} \\
            $l^{(i,j)} \gets$ \textsc{CalcNeighbors} ($l^{(i,j)}, E^{(i,j)}$)\; \label{alg1:18}
            $\mathcal{G}^{(i,j)} \gets$ \textsc{CalcMovement} ($\mathcal{G}^{(i,j)}, l^{(i,j)}$)\; \label{alg1:20}
        }
    }
    \textbf{\textit{Store}} ($\mathcal{G}_{i}$)\; \label{alg1:line21}
}

\end{algorithm}

\begin{table*}[!t] 
\centering

\caption{Related work comparison. {\bf{SMD}} = Simulated Molecular Dynamics annealing; {\bf{MC}} = Monte Carlo based method; {\bf{TCU}} = Tensor Core; {\bf{Success Rate}} = percentage of Top-10
predicted complexes.}
\resizebox{\linewidth}{!}
{
\setlength\tabcolsep{1.1pt}
    \begin{tabular}{|l| c| c| c| l| l| c| r| r| r|}
    \hline
    \textbf{Method} & \textbf{Alg.} & \textbf{Addr. Flexibility} & \textbf{Benchmark} & \textbf{System} & \textbf{Accelerator} & \textbf{TCU} & \textbf{Scaling} & \textbf{Success Rate} & \textbf{Runtime} \\
    \hline
    \hline
    PIPER~\cite{kozakov2006piper} & FFT & ✘ & BM2 & BlueGene/L & PowerPC & ✘ & 512 nodes & 9/42 (21\%) & 2 min \\
    PIPER-GPU~\cite{sukhwani2009gpu} & FFT & ✘ & BM2 & GPU & Tesla C1060 & ✘ & 1 GPU & 9/42 (21\%) & 4 h \\
    ClusPro~\cite{landaverde2014gpu} & FFT & ✘ & CAPRI Rds.13\-35 & GPU & Tesla K20C & ✘ & 1 GPU & 19/42 (45\%) & 40.8 min \\
    pyDOCK~\cite{cheng2007pydock} & FFT & ✘ & BM5.2(deduplicate) & CPU & Intel Xeon & ✘ & 32 cores & 11/55 (20\%) & 1 h \\
    MEGADOCK~\cite{ohue2014megadock} & FFT & ✘ & BM4 & TSUBAME 2.5 & Tesla K20X & ✘ & 1K GPUs & 8/176 (4\%) & 0.04 sec \\ 
    ZDOCK~\cite{pierce2011accelerating,pierce2005m,mintseris2007integrating,chen2003zdock} & FFT & ✘  & BM5(deduplicate) & CPU &-- & ✘ & -- & 15/55 (27\%) & 1.5 h \\ \hline \hline
    HADDOCK~\cite{de2010haddock} & SMD & ✔ & CASP-CAPRI & CPU & -- & ✘ & -- & 16/25 (64\%) & $\ge$100 h \\
    RosettaDock~\cite{lyskov2008rosettadock,marze2018efficient} & MC & ✔ & BM5.2 & CPU & -- & ✘ & -- & 41/88 (47\%) & 65 h \\
    SwarmDock~\cite{torchala2013swarmdock,moal2010swarmdock} & SI & ✔ & BM5.2(deduplicate) & CPU & -- & ✘ & -- & 22/55(38.18\%) & $\ge$36 h \\
    LightDock~\cite{jimenez2018lightdock,roel2020lightdock} & GSO & ✔ & BM5.2(deduplicate) & CPU & AMD Opteron & ✘ & 48 cores & 51/55 (92.7\%) & 3.3 h \\
    LightDock-Rust~\cite{roel2020integrative} & GSO & ✔ & BM5.2(deduplicate) & CPU & Intel Xeon & ✘ & 10 cores & 51/55 (92.7\%) & 1.9 h \\ \hline \hline
    {\bf\color{blue}{\method{} }(ours)} & GSO & ✔ & BM5.2(deduplicate) & GPU & Tesla A100 & ✔ & 512 GPUs & 51/55 (92.7\%) & 7 sec \\
    \hline
    \end{tabular}\label{tab:compare}
}

\end{table*}

\section{Background \& Related Work}\label{sec:background}

\subsection{Macromolecular Docking using GSO}

Macromolecular docking predicts new complexes from two known macromolecules (the {\textit{receptor}} and {\textit{ligand}}) using a sampling-and-scoring approach: (1) generating potential complex structures ({\textit{sampling}}) and (2) evaluating near-native candidates ({\textit{scoring}}). LightDock improves docking flexibility and success rates via GSO (belonging to the swarm intelligence family), with three steps: swarm initialization, glowworm agent initialization, and simulation (Alg.~\ref{alg:lightdock}).

\textbf{Swarm initialization.} This step initializes the swarm centers $I \in \mathbb{R}^{N_s\times 3}$ on the receptor surface, where each point represents a potential receptor–ligand binding site (line~\ref{alg1:line1}). It first traverses all atoms in receptor and generates the backbone (i.e. atoms of type \texttt{C} or \texttt{CA}).
The receptor surface area $Sur$ is then computed using \texttt{FreeSASA}~\cite{mitternacht2016freesasa}, and the number of swarms is set as $N_s = Sur/d$, where $d$ denotes the surface density ($\mathring{A}^2$/atom).
Finally, surface atoms are derived from backbone atoms, and K-means clustering is applied to generate swarm centers uniformly distributed around the receptor.

\textbf{Glowworm agent initialization.} This step initializes the glowworm agent vectors $\mathcal{G} \in \mathbb{R}^{N_s\times N_g\times(3+4+M_l+M_r)}$ (line~\ref{alg1:line2}), where $N_g$ is the number of agents per swarm. Each agent vector $\mathcal{G}^{(i,j)}\in \mathbb{R}^{3+4+M_l+M_r}$ of glowworm agent $(i,j)$ encodes a candidate transformation of the receptor and ligand backbones for complex formation. It comprises a translation vector $\alpha \in \mathbb{R}^{3}$, a rotation vector $\beta \in \mathbb{R}^{4}$, and an anisotropic network model (ANM) vector $\gamma \in \mathbb{R}^{M_l + M_r}$.
$\alpha$ is the movement in Euclidean space, which is randomly initialized within a sphere of radius $r$ around $I$. 
$\beta$ is the agent rotation, which is updated and initialized in four-dimensional quaternion space~\cite{shoemake1985animating}.
$\gamma$ represents the deformation magnitude along each non-trivial normal mode of the receptor and ligand, for backbone flexibility~\cite{doruker2000dynamics,atilgan2001anisotropy}.


\textbf{Simulation.} This procedure runs $S$ steps of simulation to optimize each glowworm agent according to the energy score and identifies $N_s \times N_g$ energetic minima (i.e. potential receptor-ligand complex structures). Each step includes docking pose preparation, energy score calculation, neighbor list construction, and movement. 

In docking pose preparation (lines~\ref{alg1:line7}-\ref{alg1:line12}), LightDock retrieves each atom in the receptor and ligand backbones to calculate {\textit{receptor poses}} $\mathcal{A}\in\mathbb{R}^{N_s \times N_g \times N_r\times3}$ and {\textit{ligand poses}} $\mathcal{B}\in\mathbb{R}^{N_s \times N_g\times N_l\times3}$ of $N_s\times N_g$ glowworm agents with the corresponding $\mathcal{G}$. The receptor and ligand poses denote the specific conformations and orientations transformed by $\mathcal{G}$.

In energy score calculation (lines~\ref{alg1:14}-\ref{alg1:16}), 
LightDock adopts the {\textit{distance-scaled, finite, ideal-gas reference state (DFIRE)}}~\cite{yang2008ab} to calculate the interaction energy score for each docking pose. It begins by computing the full {\textit{pairwise distances}} $\mathcal{D} \in \mathbb{R}^{N_r \times N_l}$(line~\ref{alg1:14}). For agent $(i,j)$, let $\mathcal{A}^{(i,j)}\in\mathbb{R}^{N_r\times3}$ and $\mathcal{B}^{(i,j)}\in\mathbb{R}^{N_l\times3}$ denote the transformed receptor and ligand pose. Then, $\mathcal{D}$ is denoted as
\begin{equation}\footnotesize
    \mathcal{D}_{x,y} = \sqrt{\sum_{k=1}^3(\mathcal{A}_{x,k}^{(i,j)} - \mathcal{B}_{y,k}^{(i,j)})^2}. \label{eq:dist_eq}
\end{equation}
Each distance value is then constructed into an index (binning) to retrieve its corresponding energy value from the pre-measured distance-energy lookup table $\theta$ (line~\ref{alg1:15}). Finally, the total energy score $E$ of the agent $j$ in swarm $i$ is obtained by accumulating all energy values (line~\ref{alg1:16}). 


In neighbor list construction and movement, LightDock first builds the neighbor list $l \in \mathbb{R}^{N_g}$ for each agent vector $\mathcal{G}^{(i,j)}$, containing the IDs of agents in swarm $i$ with lower energy scores (line~\ref{alg1:18}). Each agent then randomly selects a neighbor $\mathcal{G}_{i,j}'$ from $l_{i,j}$ and updates its $\alpha$, $\beta$, and $\gamma$. $\alpha$ is updated via a gradient step, $\beta$ via Spherical Linear Interpolation (SLERP)\cite{morrison1992quaternion}, and $\gamma$ via ProDy\cite{bakan2011prody}.

\begin{figure*}[t]
    \centering
    \includegraphics[width=\textwidth]{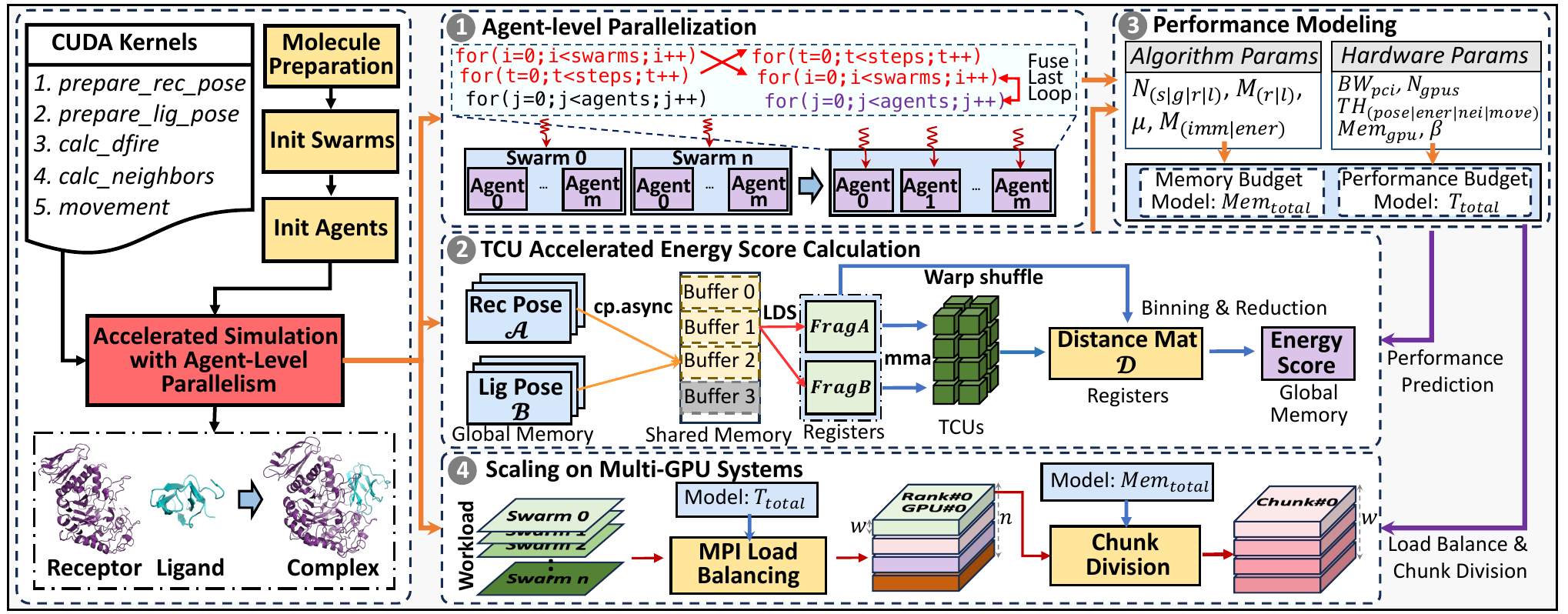}
    \caption{Overview of the \method{} framework. \method{} consists of molecule preparation, swarm initialization, agent initialization, and simulation (bottleneck). To accelerate the simulation, we devise agent-level parallelization, TCU accelerated energy score calculation, and performance model-guided scaling optimization.}     \label{fig:system_overview}
\end{figure*}

\subsection{Tensor Core Units}
Modern GPU supercomputers equipped with \textit{NVIDIA Tensor Core GPUs} provide unprecedented computational capability for scientific applications~\cite{choquette2021nvidia,dakkak2019accelerating,chen2024convstencil,liu2022toward,niu2025berrybees,li2021tcfft, 10.1145/3650200.3656634, gallet2022leveraginggputensorcores}. 
They feature specialized accelerators, \textit{Tensor Core Units (TCUs)}, designed to boost the performance of standard matrix multiplication and accumulation (MMA). In Ampere and above~\cite{9365803,10070122}, TCUs also support FP64 calculation, providing the opportunity for accelerating scientific applications. In warp level, the TCU maps the MMA into three register fragment layouts, denoted as \texttt{fragA}, \texttt{fragB}, and \texttt{fragC}. WMMA exposes these fragments through warp-level abstractions~\cite{markidis2018nvidia,fan2024dtc}, but provides limited control over register layouts and data movement. In contrast, PTX \textit{mma} offers finer-grained register-level control at a lower abstraction level. \textit{CuTe}~\cite{cecka2026cute} provides C++ abstractions for tensor layouts and MMA operations, enabling flexible thread--data mappings and register configurations without directly managing low-level PTX instructions.




While TCUs offer substantial acceleration opportunities, they demand structured computation patterns tailored to specific matrix operations. Given the diversity of the computing workloads, this specialization is often poorly aligned with many scientific applications. As a result, this mismatch limits the achieved peak performance and practical scalability of scientific computing on TCUs. Recent works have explored using TCUs beyond the standard MatMul, such as reductions and scans~\cite{dakkak2019accelerating}, stencil computations~\cite{chen2024convstencil,liu2022toward}, Breadth-First Search (BFS)\cite{niu2025berrybees}, FFT\cite{li2021tcfft}, and scientific computing applications with irregular computational patterns~\cite{10.1145/3712285.3759829}. 
In LightDock, however, the dominant bottleneck lies in the more complicated pairwise distance calculation for docking energy scoring, which requires computing distances between every atom pair across the docking macromolecules. 
However, TCUs require highly structured matrix computations, which are fundamentally mismatched with the irregular pairwise interaction patterns in GSO-based docking. 
This mismatch necessitates careful reformulation to enable the efficient utilization of TCU.

\subsection{Related Work}

\textbf{Rigid body docking.}
FFT-based rigid-body docking fixes one molecule and samples the other on a grid to evaluate interaction energies~\cite{desta2020performance}. 
This formulation enables efficient high-throughput screening. 
Representative methods include FTDock~\cite{gabb1997modelling}, ZDOCK~\cite{chen2003zdock}, and PIPER~\cite{kozakov2006piper}, which improve accuracy through enhanced scoring functions while retaining high computational efficiency.


\textbf{Flexible docking.}
Flexible docking accounts for conformational changes upon binding, achieving higher accuracy at the cost of significantly increased computation. 
Methods such as SwarmDOCK~\cite{moal2010swarmdock}, HADDOCK~\cite{dominguez2003haddock}, and RosettaDock~\cite{marze2018efficient} employ stochastic optimization or sampling-based strategies. 
LightDock~\cite{jimenez2018lightdock,roel2020lightdock,jimenez2023lightdock} further improves robustness using Glowworm Swarm Optimization (GSO) to achieve high docking accuracy across diverse macromolecules, enabling robust docking across proteins, peptides, and DNA.

\textbf{Performance and accuracy.} 
Tab.~\ref{tab:compare} compares representative macromolecular docking methods. Rigid-body docking is computationally efficient and often accelerated on GPUs or many-core CPUs. For example, PIPER and MEGADOCK complete docking in 2 minutes and 0.04 seconds on the BlueGene/L~\cite{adiga2002overview} and TSUBAME 2.5~\cite{matsuoka2017tsubame2} supercomputers, enabling high-throughput screening. In contrast, flexible docking provides higher accuracy (e.g. over 47\% Top-10 success rate) but incurs orders-of-magnitude higher computational cost, often requiring hours to days per docking. For example, HADDOCK may take hundreds of hours, while LightDock requires several hours even with parallel execution~\cite{roel2020integrative}. \method{} bridges this gap, enabling nearly real-time flexible docking on GPU-accelerated supercomputers.

\section{Design of SparkleDock}\label{sec:design}

\begin{algorithm}[!t] 
\small
\caption{Agent-level parallelization (same input \& output as Alg.~\ref{alg:lightdock})} \label{alg:opt_design}
\SetKwInOut{Input}{Input}
\SetKwInOut{Output}{Output}
\SetKwProg{KwKernel}{\_\_global\_\_ void}{}{end}
\SetKwFunction{Launch}{Launch}{}
\SetKwComment{Comments}{ $\triangleright$}{}


\SetKwFor{ForPar}{for}{do in parallel}{endfor}

{\color{blue}\tcp*[h]{Swarm and glowworms initialization}}\\
\For{$t \gets 0$ \KwTo $S-1$}{ \label{alg_agent:line_s} 
{\color{blue}\tcp*[h]{Agent parallelism with five CUDA kernels}}\\
\ForPar{$(i,j) \gets (0,0)$ \KwTo $(N_s-1,N_g-1)$}{ \label{alg_agent:line_cuda_start} 
    {\color{blue}\tcp*[h]{--- 3D grid computations ---}}\\
    $\mathcal{A}^{(i,j)} \gets$ $<$\textsc{Grid3D}$>$ \textit{prepare\_rec\_pose}($\mathcal{R}, \mathcal{G}^{(i,j)}$) \;
    $\mathcal{B}^{(i,j)} \gets$ $<$\textsc{Grid3D}$>$ \textit{prepare\_lig\_pose}($\mathcal{L}, \mathcal{G}^{(i,j)}$) \;
    $E^{(i,j)} \gets$  $<$\textsc{Grid3D}$>$ \textit{calc\_dfire}($\mathcal{A}^{(i,j)}, \mathcal{B}^{(i,j)}$) \;

    {\color{blue}\tcp*[h]{--- 1D grid computations ---}}\\
    $l^{(i,j)} \gets$  $<$\textsc{Grid1D}$>$ \textit{calc\_neighbors}($l^{(i,j)},E^{(i,j)}$) \;
    $\mathcal{G}^{(i,j)} \gets$ $<$\textsc{Grid1D}$>$ \textit{movement}($\mathcal{G}^{(i,j)}, l^{(i,j)}$) \; \label{alg_agent:line_cuda_end} 
}
}
\textbf{\textit{Store}} ($\mathcal{G}$) \;
\end{algorithm}

Fig.~\ref{fig:system_overview} illustrates the design of the \method{} framework, including four stages: (1)~Molecule preparation for data loading and pre-processing. (2)~Swarm initialization. (3)~Glowworm agent initialization. (4)~Accelerated simulation with agent-level parallelism.
\textbf{Among them, the simulation stage accounts for over 99\% of the total runtime.}
To leverage the computational power of multi-GPU systems, we provide several optimizations for \method{}. 
We first devise a fine-grained, agent-level parallelization algorithm (step-1 in Fig.~\ref{fig:system_overview}) to enhance the parallelism and exploit the compute throughput of the individual GPU. Next, we reformulate the hotspot energy score calculation (over 89\% of the total runtime) to exploit the TCU acceleration and also introduce optimized CUDA kernels that do partial computation that results from the reformulation. Based on the optimized designs and hardware parameters measured by microbenchmarks, we construct a memory budget model, $Mem_{total}$, and a performance budget model, $T_{total}$, to predict memory footprint and docking performance. Finally, guided by $Mem_{total}$ and $T_{total}$, we propose MPI load balancing and out-of-core chunk division strategies to improve the scalability on multi-GPU systems.

\subsection{Fine-grained Agent-level Parallelism} \label{sec:fine-grained-par}
The agent vectors $\mathcal{G}$ at each step are updated from the previous step, so the $N_s$ swarm loops and the $N_g$ agent loops can be unrolled with each agent executing the GSO simulation independently. 
We exploit the parallelization opportunity to perform loop reordering (step-1 in Fig.~\ref{fig:system_overview}) and introduce agent-level parallelism in Alg.~\ref{alg:opt_design} to accelerate the time-consuming simulation. As shown in Fig.~\ref{fig:system_overview}, we first perform loop reordering between the swarm and simulation loops. We then merge and unroll the swarm and agent loops, mapping them onto CUDA threads to derive the Alg~\ref{alg:opt_design}.
This algorithm begins with a simulation loop that executes $S$ GSO steps (line~\ref{alg_agent:line_s}). For each GSO step, we devise five efficient CUDA kernels with dedicated thread blocks to execute the agent-level GSO optimization(line \ref{alg_agent:line_cuda_start}-\ref{alg_agent:line_cuda_end} in Alg.\ref{alg:opt_design}). Docking pose preparation and DFIRE score calculation exhibit high parallelism at the atom level. Therefore, we map these calculations onto a 3D grid with 1D block. The neighbor calculation and movement exhibit limited scalability at the glowworm level. We use the 1D grid with 1D block to unroll the swarm and glowworm loops.
Compared to swarm-level parallelization, it increases parallelism by more than $N_g$ times. 

\subsection{Restructuring Energy Scoring into
TCU Formulation}
\label{sec:cal_energy}

The energy score calculation is the hotspot of the docking simulation, occupying over 89\% of the total execution time. 
We developed a TCU-accelerated CUDA kernel using \textit{CuTe} from \textit{CUTLASS}. 


\begin{figure}[!t]
    \centering
    \includegraphics[width=0.9\columnwidth]{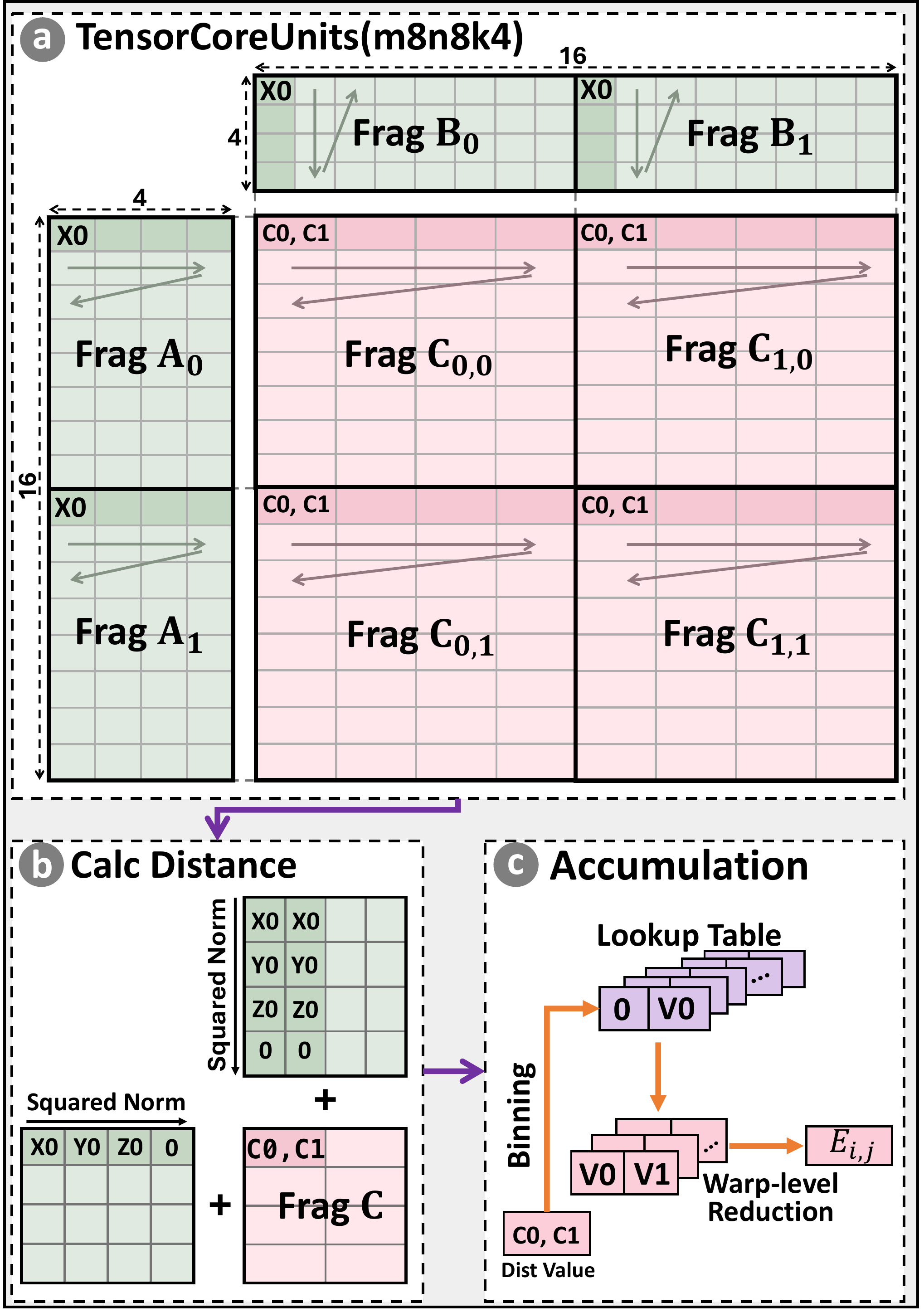}
    \caption{Implementation of the TCU-accelerated \textit{calc\_dfire} kernel. (a) TCU-aware warp-level partition. (b) Thread-level distance calculation. (c) Accumulate the energy scores.}
    \label{fig:pairwise_dist}
\end{figure}

\textbf{Restructuring energy scoring for TCUs.} 
Based on the pairwise distance calculation in Eqn~\ref{eq:dist_eq}, we expand it as follows:
\begin{equation}\footnotesize
\begin{aligned}
    \mathcal{D}_{x,y} = \sqrt{\sum_{k=1}^3{(\mathcal{A}_{x,k}^{(i,j)})^2} + \sum_{k=1}^3{(\mathcal{B}_{y,k}^{(i,j)})^2} - \underline{2 \cdot {\sum_{k=1}^3\mathcal{A}_{x,k}^{(i,j)} \cdot \mathcal{B}_{y,k}^{(i,j)}}}},
    \label{eq:dist}
\end{aligned}
\end{equation}
where the underlined term is the tall-skinny matrices multiplication, and the remaining terms are the squared norm(i.e. sums of squares) for matrices $\mathcal{A}$ and $\mathcal{B}$.
Based on the derivation, the pairwise distance calculation can be decomposed into three steps:

(1)~Matrix multiplication $\mathcal{C} = \mathcal{A}\times\mathcal{B}^T$ adapted for TCU.

(2)~Calculate Row-wise and column-wise squared norms of $\mathcal{A}$ and $\mathcal{B}$.

(3)~Element-wise combination of these terms to construct the distance matrix.

Note that in this formulation, (1) and (2) are data-independent, enabling simultaneously execution of (1) on TCUs and (2) on CUDA cores~\cite{liu2024juno}.



\textbf{TCU kernel implementation.} 
Based on Eqn.~\ref{eq:dist}, we reformulate the energy score calculation into a TCU-accelerated kernel called \textit{calc\_dfire}. Fig.~\ref{fig:pairwise_dist} shows the design of the TCU kernel in warp level.


Based on the agent-level parallelization, we use a 3D grid with 1D blocks (128 threads per block), where the \textit{gridDim.z} unrolls the swarm loop $N_s$, the \textit{gridDim.y} unrolls the glowworm agent loop $N_g$, and the \textit{gridDim.x} tiles the data block of matrix $\mathcal{A}$ and matrix $\mathcal{B}$. Each block is decomposed into four warps to perform the TCU-accelerated \textit{calc\_dfire} kernel using \textit{mma} simultaneously. Each warp then launches four \textit{mma} instructions to compute a $16 \times 16$ tile, improving SM occupancy. Fig.~\ref{fig:pairwise_dist}(a) illustrates the specific partition in warp level. The \textbf{\textit{k}} dimension of the receptor pose $\mathcal{A}$ and ligand pose $\mathcal{B}$ is 3, which is lower than the minimum dimension of the \textit{m8n8k4} fragment in FP64 \textit{mma} (i.e. \textbf{\textit{k}}=4). Therefore, we allocate $16 \times 4$ shared memory tiles for both $\mathcal{A}$ and $\mathcal{B}$, and employ a strided memory copy to pad zeros along the \textbf{\textit{k}} dimension to 4.

Following the above partition, we propose a monolithic design to consolidate the entire energy score calculation into a single CUDA kernel. Based on the derivation in Eqn.~\ref{eq:dist}, we first leverage the \textit{FP64} \textit{mma} instruction, allowing each warp to execute the \textit{m8n8k4} matrix multiplication. 
TCUs and CUDA cores share the same register files. During the \textit{mma} calculation, each thread can directly load the relevant elements from its allocated register fragments \texttt{fragA} and \texttt{fragB} to compute the squared norms, and subsequently accesses its accumulated results in accumulator \texttt{fragC} to calculate the pairwise distances $\mathcal{D}$. 
Using \texttt{get\_slice(threadIdx.x)} in \textit{CuTe} allows each thread to obtain its assigned registers in \textit{mma}. Finally, the distance values are discretized into indices through \textit{DistBinning} in Alg.~\ref{alg:lightdock}. These indices are subsequently employed to query the lookup table $\theta$, yielding the DFIRE energy contributions. These values are finally accumulated to obtain the final DFIRE energy score of each agent. We devise an efficient warp-level reduction using shared memory to accumulate the final energy score. Based on our design, all computations within the \textit{calc\_dfire} are performed on-chip, avoiding redundant data movements, and improving SM utilization.

\begin{figure}[!t]
    \centering
    \includegraphics[width=\linewidth]{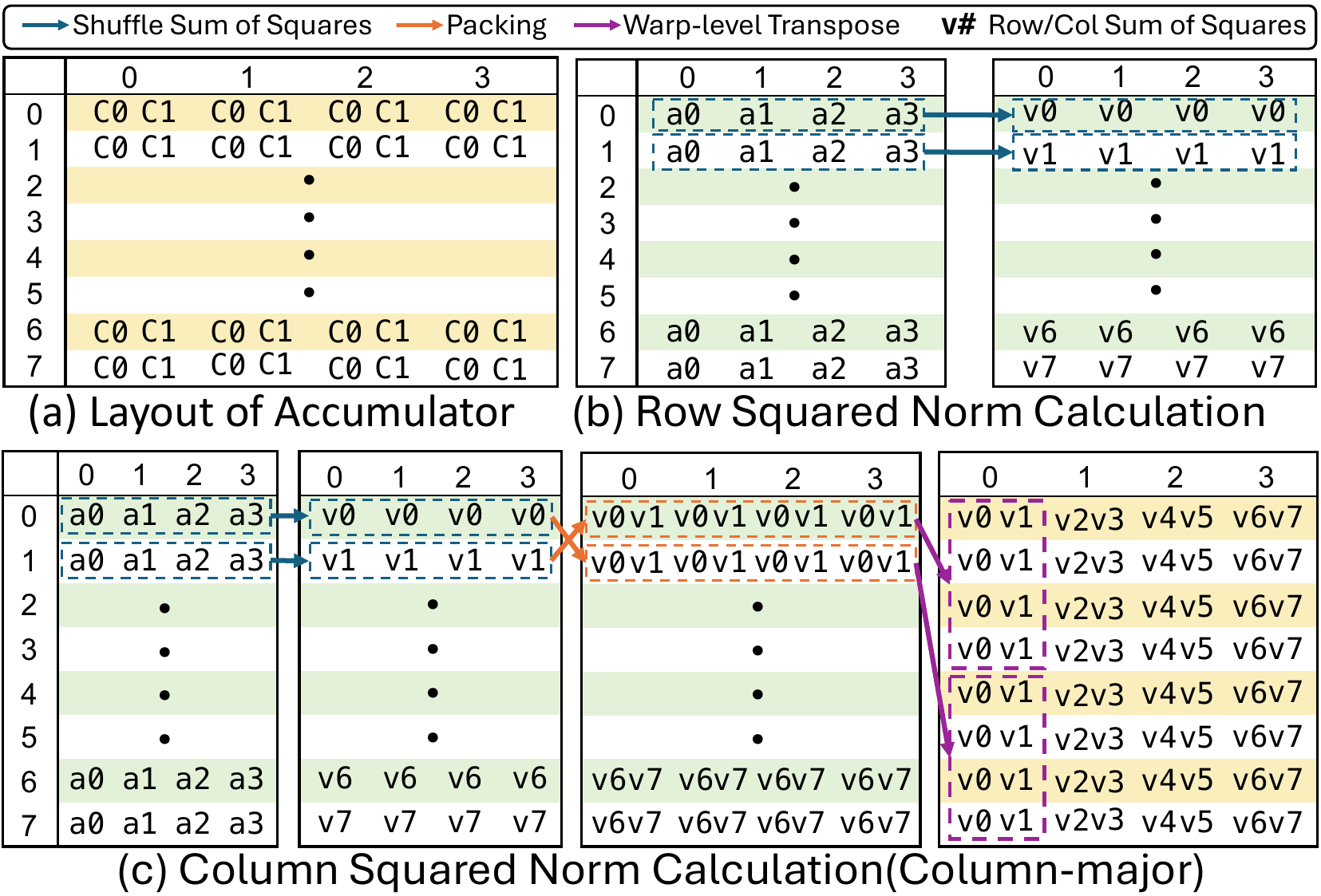}
    \caption{Register remapping for sum of squares calculation. (a) The \textit{\textbf{mma.m8n8k4}} register layout. (b) Register remapping for fragment $\mathcal{A}$. (c) Register remapping for fragment $\mathcal{B}$.}
    \label{fig:remapping}
\end{figure}


\lstset{
  language = C++, 
  basicstyle=\ttfamily\small, 
  keywordstyle = {\bfseries \color[cmyk]{0,1,0,0}}, 
  commentstyle = {\itshape \color[cmyk]{1,0.4,1,0}}, 
  stringstyle = {\ttfamily \color[rgb]{0,0,1}},
  numbers=left,
  numberstyle=\tiny,
  numbersep=3pt, 
  breaklines=true, 
  breakindent = 0pt, 
  lineskip={-6pt},
  columns=flexible, 
  keepspaces=true,
  breaklines=true,  
  xleftmargin=0pt,  
  xrightmargin=5pt,  
  showspaces=false,                
  showstringspaces=false,
  showtabs=false,    
  language=C++,               
  showstringspaces=false,
  showtabs=false,  
  frame=tb, 
  morekeywords={__device__, threadIdx, T},
  emph={__shfl_down_sync, __shfl_up_sync, __shfl_sync, __shufl_sync, typename}, emphstyle={\color{blue}},
}
\lstset{escapeinside={<@}{@>}}

\begin{figure}[t]
\centering
\begin{lstlisting}[caption = {Design of the squared norm calculations.
}, label = list:sum_squares]
template<typename T>
__device__ T <@\textbf{shufl\_sum\_squares}@>(T src, int K){
  T sum = 0;
  for i = 0 to K-1
    T s = _shufl_sync(MASK, src, i, K);
    sum += s * s;
  return sum;
}
 
template<typename T>
__device__ <@\textbf{col\_squared\_norm}@>(T dst[2],T src,int K){
  int lane = threadIdx.x % 32;
  int row = lane % 8;
  int col = lane / 8;
  int idx = row * 8 + col;
  T sum = <@\textbf{shufl\_sum\_squares}@>(src, K);
  // Column packing
  dst[0] = _shfl_up_sync(MASK,sum,4,8);
  dst[1] = _shfl_down_sync(MASK,sum,4,8);
  // Warp-level transpose
  T m = _shfl_sync(MASK,dst[0],idx,32);
  dst[0] = m;
  m = _shfl_sync(MASK,dst[1],idx,32);
  dst[1] = m;
}
\end{lstlisting} 
\end{figure}


\textbf{Efficient register remapping for TCU.}
In FP64 \textit{mma} operation(Fig.~\ref{fig:pairwise_dist}(a)), \texttt{fragA} follows a row-major layout and \texttt{fragB} follows a column-major layout, where each thread holds one element (i.e. register). The accumulator \texttt{fragC} is organized in a row-major layout, where each thread holds two elements(i.e. Fig.~\ref{fig:remapping}(a)).
These heterogeneous layout designs introduce mismatches across fragments, leading to additional data movement when computing the squared norm.
We devise a lightweight register-level remapping strategy based on warp shuffle to replace these redundant data movements. Through \emph{micro-benchmark}~\cite{abdelkhalik2022demystifying} evaluation on the A100 GPU, loading a 64-bit value from shared memory to a register takes approximately 23 cycles, whereas a warp shuffle takes only 2 cycles. This motivates our design to perform squared norm computation entirely at the register level, avoiding costly memory accesses.






Fig.~\ref{fig:remapping}(b) shows the register remapping for \texttt{fragA}. We propose a \texttt{shuffle\_sum\_of\_squares()} method to efficiently compute the row squared norm of \textit{A}, as shown in List.~\ref{list:sum_squares}. In this algorithm, \texttt{\_\_shufl\_sync(i,k)} is used to broadcast the value of column \texttt{i} to register \texttt{s} of all threads within the same row (i.e. line~5 of List.~\ref{list:sum_squares}). Each thread then calculates the squared values of \texttt{s} and accumulates in \texttt{sum}. After the above method, each thread holds the row sum of squares, which can be directly added to \textit{C} in element-wise.

For \texttt{fragB}, the column-major layout introduces additional mismatch with \texttt{fragC}. Since each thread corresponds to two column elements in \texttt{fragC}, we design a column-wise squared norm computation with a register packing strategy in Fig.~\ref{fig:remapping}(c). Specifically, we design a \texttt{col\_squared\_norm} procedure in List.~\ref{list:sum_squares}.
This function first invokes \texttt{shuffle\_sum\_of\_squares()} to compute the column-wise sum of squares for each thread (line~16). It then applies a packing approach to combine two column registers into one thread. 
Specifically, \texttt{\_\_shufl\_up\_sync(4,8)} shifts the even columns into the odd, while the \texttt{\_\_shufl\_down\_sync(4,8)} shifts the odd columns into the even columns.
After packing, each thread holds two column registers, forming a column-major $8 \times 4$ squared norm matrix. We perform a warp-level transpose operation (lines~20-24) to convert it into an $8 \times 4$ row-major squared norm matrix matching the \texttt{fragC} layout. Finally, the accumulator \texttt{fragC} performs an element-wise addition of the two squared norm matrices to obtain the $\mathcal{D}$.

Each thread loads one row of $\mathcal{A}$(four elements) and two columns of $\mathcal{B}$(eight elements) from shared memory to \texttt{fragA} and \texttt{fragB} and computes the sum of squares in the intuitive design, consuming around \textbf{276} cycles. However, only \textbf{12} warp shuffle instructions are required in register remapping with \textbf{24} cycles. This optimization theoretically accelerates by 12$\times$.



\begin{figure}[!t]
    \centering
    \includegraphics[width=\columnwidth]{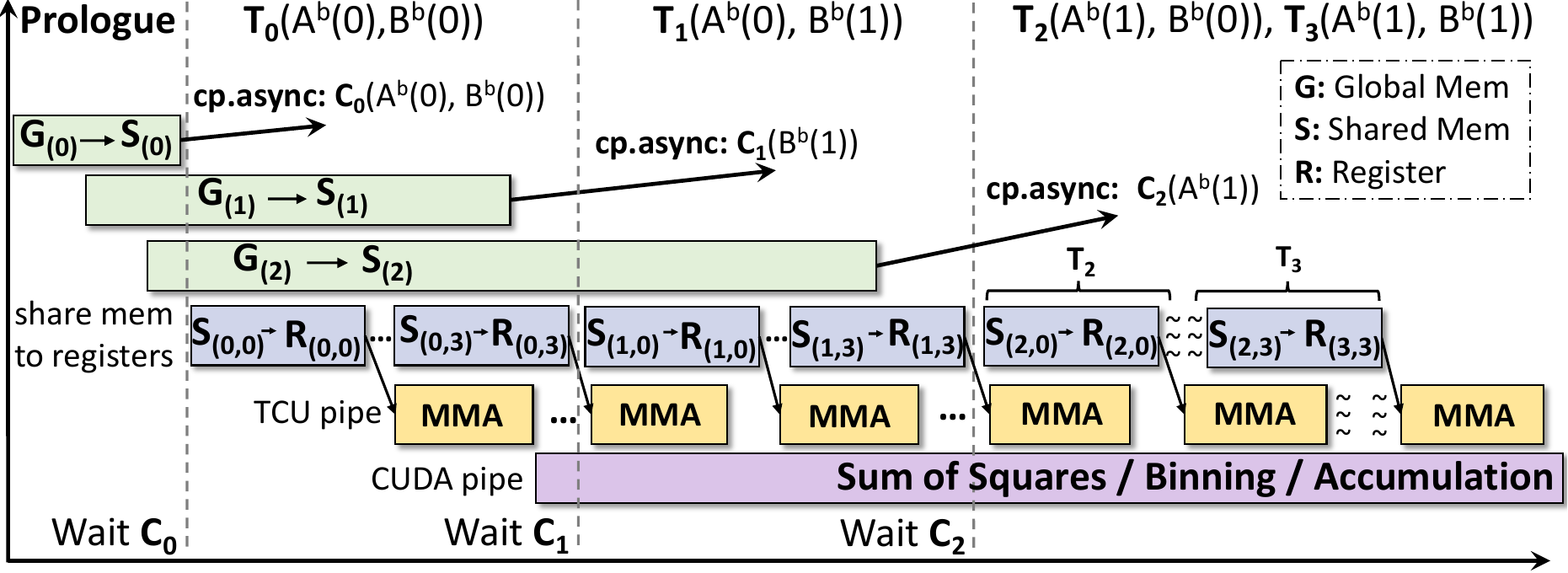}
    \caption{Overlapping the memory and calculation pipeline using \textbf{\textit{cp.async}}. Each CUDA thread block is extended into four pipeline stages, where each stage handles a 32$\times$32 tile.}
    \label{fig:overlap}
\end{figure}

\textbf{Pipeline optimization.} 
On Ampere and later GPUs, the \textit{cp.async} instruction enables direct asynchronous copying from global to shared memory, bypassing the register file. This mechanism eliminates the register-level dependencies and enables multi-stage asynchronous copy pipelines with explicit synchronization. We exploit this capability to overlap memory transfers with TCU computation, as
illustrated in Fig.~\ref{fig:overlap}.

Following the above warp-level partitioning, each CUDA thread block processes a $64\times64$ tile of the pairwise-distance matrix, which is divided into four $32\times32$ pipeline buffers. Let
$\mathcal{A}^{b},\mathcal{B}^{b}
\in\mathbb{R}^{64\times3}$ denote the corresponding receptor and
ligand coordinate tiles executed by the CUDA thread block. Then, the four pipelines are denoted as $T_0$($\mathcal{A}^b(0)$,$\mathcal{B}^b(0)$), $T_1$($\mathcal{A}^b(0)$,$\mathcal{B}^b(1)$), $T_2$($\mathcal{A}^b(1)$,$\mathcal{B}^b(0)$), and $T_3$($\mathcal{A}^b(1)$,$\mathcal{B}^b(1)$), where $A^b(i), B^b(i)\in\mathbb{R}^{32\times 3}$ are receptor and
ligand sub-matrices in each pipeline. Within each execution pipeline, the CUDA block further computes on four $16 \times 16$ subtiles in Fig.~\ref{fig:pairwise_dist}(a) (i.e. four warps per block).
To supply the four compute stages, we organize the required data transfers into three asynchronous-copy groups: $C_0$($\mathcal{A}^b(0)$,$\mathcal{B}^b(0)$), $C_1$($\mathcal{B}^b(1)$), $C_2$($\mathcal{A}^b(1)$). The first $C_0$($\mathcal{A}^b(0)$,$\mathcal{B}^b(0)$) load $\mathcal{A}^b(0)$ and $\mathcal{B}^b(0)$ from global memory to shared memory. After $C_0$($\mathcal{A}^b(0)$,$\mathcal{B}^b(0)$) is finished, $S_0$ begins the computation while the second group, $C_1$($\mathcal{B}^b(1)$), asynchronously loads $\mathcal{B}^b(1)$ to the shared memory. When $C_1$($\mathcal{B}^b(1)$) completes, $S_1$ and cp($\mathcal{A}^b(1)$) are issued concurrently. After cp($\mathcal{A}^b(1)$) completes, the remaining stages ($S_2$, $S_3$) can proceed. The $S_3$ pipeline calculation requires $\mathcal{A}^b(1)$ and $\mathcal{B}^b(1)$, both of which are already resident in shared
memory, thereby avoiding additional data transfers.
Note that four pipeline buffers are used due to the limited shared memory capacity per SM.

\begin{table}[]
\caption{Performance model parameters for \method{}.}
\resizebox{\linewidth}{!}
{
\setlength\tabcolsep{0.9pt} 
\begin{tabular}{|c|l|l|}
\hline
\multicolumn{1}{|l|}{} &  \textbf{Description} & \textbf{Parameters} \\ \hline \hline
\multirow{5}{*}{\rotatebox{90}{Algorithm}}    & Swarm/agent num & $N_{[s/g]}$  \\ \cline{2-3} 
& Receptor/ligand backbone atoms num & $N_{[r/l]}$\\ \cline{2-3} 
&ANM vector size & $M_{[l/r]}$ \\ \cline{2-3} 
& lookup table / intermediate variables size  & $M_{[ener/imm]}$ \\ \cline{2-3} 
& Moving mask&$\mu$ \\ \hline
\multirow{4}{*}{\rotatebox{90}{Hardware}}      & Throughput of CUDA kernels & $TH_{[ener/nei/move/pose]}$  \\ \cline{2-3} 
                       & Bandwidth & $BW_{pci}$  \\ \cline{2-3} 
                       & Number of the GPUs & $N_{gpus}$ \\ \cline{2-3} 
                       & Memory footprint of a double & $b_{fp}$ \\ \hline
\end{tabular}\label{tab:parameters}
}
\end{table}

\subsection{Performance Model}
We build a performance model~\cite{snavely2002framework} to characterize both latency and memory usage of \method{}, with the aim of identifying the scaling constraints. The definitions of all parameters are summarized in Tab.~\ref{tab:parameters}.

\textbf{Memory footprint modeling.}
The memory footprint of the framework comprises four components: docking poses, glowworm agent vectors, lookup table with the array length of $M_{ener}$, and intermediate variables with array length of $M_{imm}$. The memory footprint of docking poses is denoted as 
\begin{equation}\small
    Mem_{pose} = 3\cdot b_{fp}\cdot N_s\cdot N_g \cdot (N_r + N_l),
\end{equation}
while the agent vectors are
\begin{equation} \small
    Mem_{g} = b_{fp} \cdot N_s\cdot N_g \cdot (7+M_l+M_r).
\end{equation}
Combining these, the total GPU memory budget is
\begin{equation} \small
    Mem_{total} = Mem_{pose} + Mem_{g} + b_{fp}\cdot(M_{imm} +  M_{ener}).
\end{equation}
Here, $M_{imm}$ and $M_{ener}$ are constant values. Therefore, the $Mem_{total}$ provides a closed-form estimate of the memory budget given to a specific docking task.

\textbf{Runtime modeling.}
Benefiting from agent-level parallelism, the maximum parallel degree is $N_s\times N_g$. For each agent, we divide the framework into four modules to construct the performance budget model $T_{total}$:

{(1)} DFIRE score calculation (bottleneck): The projected runtime for agent $g$ is
\begin{equation}\small
    T_{ener}^g=({8\cdot N_l \cdot N_r}/{ TH_{ener}} + {\lceil \log_2(N_l \cdot N_r)\rceil}/{TH_{ener}})\cdot \mu, \label{eq:DIFIR_perf}
\end{equation}
where $8\cdot N_l \cdot N_r$ represents pairwise distance computation, binning and indexing, the $\lceil \log_2(N_l \cdot N_r)\rceil$ represents parallel reduction in energy scoring. The binary mask $\mu \in \{0,1\}$ randomly indicates whether agent $g$ moves at step $t$: if $\mu=0$, the agent stays and its energy score is unchanged.

{(2)} Docking pose preparation: It express the total runtime of \textit{prepare\_rec\_pose} and \textit{prepare\_lig\_pose} kernels for each agent $g$, denoted as
\begin{equation} \small
    T_{pose}^g=({3\cdot (N_l+N_r)\cdot(7+M_l+M_r)})/{TH_{pose}},
\end{equation}

{(3)} Agent neighbor construction and movement: This module constructs the total runtime of \textit{calc\_neighbors} and \textit{movement} for each agent $g$, denoted as 
\begin{equation} \small
    T_{other}^g={N_g\cdot M_{nei}/{TH_{nei}} + M_{move}/{TH_{move}}},
\end{equation}
where $M_{nei}$ denotes the computational cost of constructing the neighbor list and $M_{move}$ denotes the computational cost of movement update.

{(4)} Data transfer: This module quantifies CPU–GPU communication cost, denoted as 
\begin{equation} \small
    T_{data} = ({3\cdot b_{fp}\cdot(N_r+N_l)+Mem_{g}})/{BW_{pci}},
\end{equation}
where the first term is the projected data movement time (receptor, ligand, and agent vector) from CPU to GPU, and the second term is the projected data movement time from GPU to CPU.

Combining the above components, the total $S$ steps GSO runtime $T^p_i$ of swarm $i$ can be denoted as
\begin{equation} \small
T^p_i = \sum_g^{N_g-1}\sum_{t}^{S-1}(T_{pose}^g+T_{ener}^g+T_{other}^g). \label{eq:Lp}
\end{equation}
MPI can unroll the swarm loops in parallel. Therefore, the overall model in the multi-GPU environment is
\begin{equation}\small
\begin{aligned}
    T_{total} = \max_{0\leq p< N_{gpus}}\left(\sum_{i=0}^{N_s^p-1} T^p_i\right)+T_{data},
\end{aligned}
\end{equation}
where $N_s^p$ is the swarm size assigned by MPI rank $p$.
In $T_{total}$, $TH_{[\cdot]}$ and $BW_{pci}$ can be obtained through micro-benchmarks, while $\mu$ of each step can be expressed by the generated docking random seed. Thus, we can project the total docking runtime using $T_{total}$ given to the workload of each MPI rank.

\textbf{Insights from the performance model.}
We list observations from the performance model analysis. 
\emph{(1) Performance prediction:} $T_{total}$ can predict the runtime before docking, enabling users to plan computing resources effectively.
\emph{(2) MPI load balancing:} Based on Eqn~\ref{eq:Lp}, we can devise a load balanced workload division strategy for each MPI rank.
\emph{(3) Memory chunk division:} $Mem_{total}$ can effectively express the total memory footprint of each MPI rank. For large-scale docking tasks with massive memory demand, $Mem_{total}$ can guide the automatic out-of-core chunk division of each MPI docking workload to handle larger tasks within a limited GPU memory.

\begin{algorithm}[!t]
\small
\caption{Performance Model Guided Static Load Balancing Algorithm}
\SetKwInOut{Input}{Input}
\SetKwInOut{Output}{Output}
\SetKw{KwAnd}{and}
\SetKwFor{ForPar}{for}{do in parallel}{endfor}
\SetKwComment{Comments}{ $\triangleright$ }{}
\label{alg:load_balance}
\Input{
Workload queue $Q[N_{gpus}]$; agent vectors $\mathcal{G}$
}
\Output{
$Q[N_{gpus}]$
}

\For{$p \gets 0$ \KwTo $N_{gpus}-1$}{\label{alg:balance:line1}
$Q[p].t \gets 0$;\, $Q[p].idx \gets \emptyset$

\label{alg:balance:line4}
}

\For{$i \gets 0$ \KwTo$ N_s-1$}{

    $p^\star \gets -1$; \,$T_{min} \gets \infty$   \label{alg:balance:line5}

    \For{$p \gets 0$ \KwTo $N_{gpus}-1$}{
        \If{$Q[p].t < T_{min}$}{
            $T_{min} \gets Q[p].t$; \, $p^\star \gets p$;
            
             \label{alg:balance:line10}
        }
    }

    $Q[p^\star].t \gets Q[p^\star].t + T_i^{p^\star}$ \label{alg:balance:line11}
    \Comments{\textcolor{blue} {Update via {Eqn~\ref{eq:Lp}}}} 
    
    $enqueue(Q[p^\star].idx, i)$ \label{alg:balance:line12}
}

\end{algorithm}

\subsection{Scaling \method{} on multi-GPU Platform}

Based on the design on single GPU and the performance model, we further design an MPI parallelization scheme to tile the swarm loop, enabling multi-GPU acceleration.


\textbf{Static MPI load balancing.} \label{sec:load}
Guided by the performance model $T_i^p$ in Eqn~\ref{eq:Lp}, 
we implement the MPI setup function in Alg.~\ref{alg:load_balance} to statically distribute the total workload among $N_{gpus}$ MPI ranks.
This function first initializes the computing workload $Q[N_{gpus}]$ of each MPI rank, including the total latency $Q.t$, and swarm task index queue $Q.idx$~(line~\ref{alg:balance:line1}-\ref{alg:balance:line4}). 
For each swarm $i$, the $Q[p^\star]$ with the minimal current runtime is selected (line~\ref{alg:balance:line5}-\ref{alg:balance:line10}). The execution time of task 
$i$ is then estimated via Eq.~\ref{eq:Lp} and added to $Q[p^\star]$ (line \ref{alg:balance:line11}), while the task index $i$ is appended to $Q[p^\star].idx$ (line \ref{alg:balance:line12}). After this setup, each MPI rank $p$ fetches tasks from its own task queue $Q[p].idx$, achieving balanced workload distribution. Since the \method{} involves an embarrassingly parallel in swarm loop, this setup can be seamlessly integrated into the docking initialization process to achieve a static workload dispatch before simulation.

\begin{figure}[!t]
    \centering
    \includegraphics[width=\columnwidth]{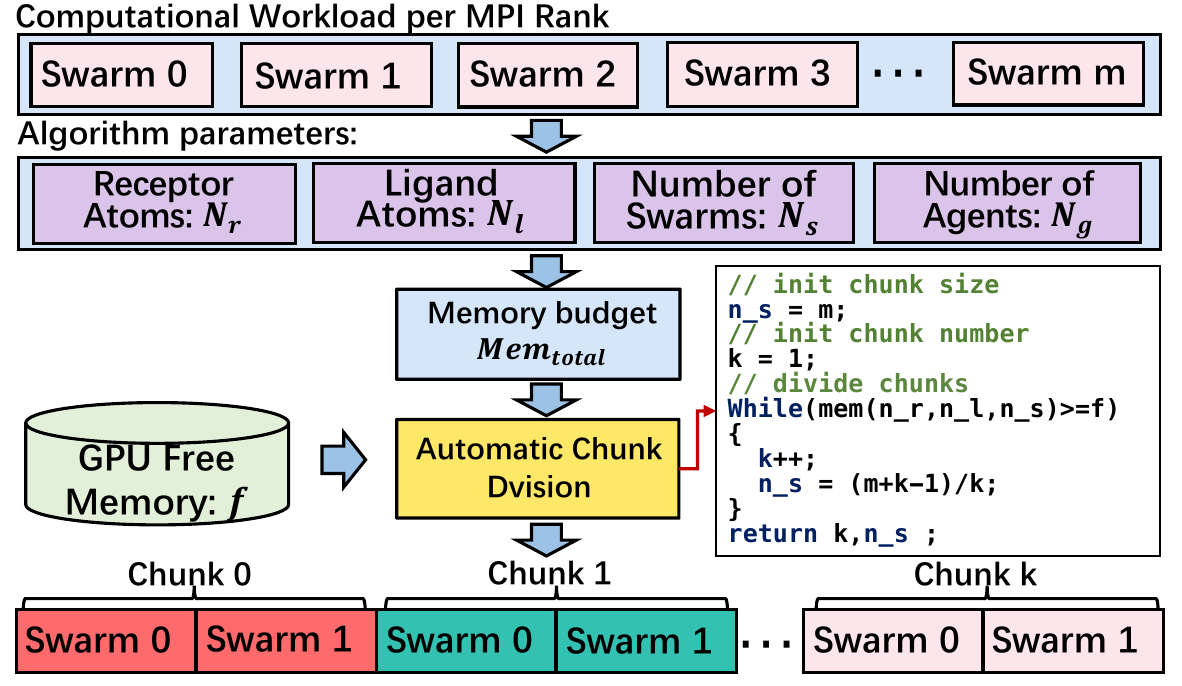}
    \caption{Performance model guided out-of-core automatic chunk division.}
    \label{fig:chunks}
\end{figure}

\textbf{Automatic out-of-core chunk division.} 
In large-scale docking tasks, each MPI consumes substantial GPU memory, often leading to out-of-memory (OOM) errors. 
We devise an automatic chunk division method to enable \method{} to adapt to various GPU memory budgets on multi-GPU environments using $Mem_{total}$. 
As shown in Fig.~\ref{fig:chunks}, for each MPI rank, we extract the algorithm parameters to construct $Mem_{total}$. Then, we retrieve the available GPU memory $f$ using the \textit{cudaMemGetInfo}. Based on $Mem_{total}$ and $f$, an automatic chunk division approach is designed to iteratively divide the MPI workload into multiple chunks until the memory requirement of the single chunk is less than the available GPU memory. 
Note that the chunk division is automatic. Once the chunk size is determined, \method{} uses the memory pool to manage GPU memory and execute the docking process for each chunk.

\section{Evaluation}\label{sec:evaluation}

\subsection{Experimental Setup}

\begin{table}[]

\caption{Complexes used in the evaluation.}
\renewcommand{\arraystretch}{1.2}
\resizebox{\linewidth}{!}
{
\setlength\tabcolsep{0.8pt} 
    \begin{tabular}{|c|c|c|c|c|c|c|c|c|c|c|}
    
    \hline
    \textbf{Dataset} & {2VXT} & {3VLB} & {2A1A} & {2GTP} &{2X9A}&{1RKE} &{4GAM}& {4JCV} &{4LW4} \\\hline \hline
    \textbf{Category} & A & EI &ES &OG & OR & OX & ER & OX & ES \\\hline
    \textbf{Swarms Num.} & 402 & 336 & 271 & 298 & 130 & 294 & 1245 & 695 & 558 \\ \hline
    \textbf{Rec Atoms Num.} & 3002 & 3052 & 2039 & 2516 & 757 & 2030 & 17307 & 6032 & 6058 \\ \hline
    \textbf{Lig Atoms Num.} & 1274 & 1658 & 1440 & 1061 & 481 & 1256 & 1125 & 1766 & 1138 \\ \hline
\end{tabular} \label{tab:dataset}
}    
\label{tab:dataset}
\end{table}

\begin{figure*}[!t]
\centering
\begin{minipage}{0.645\linewidth}
\centering

\captionof{table}{Summary of speedup. The baseline on CPU runs LightDock-Rust uses 40 threads; A100 and H100 GPU runs the proposed \method{}. } \label{tab:summary}
\resizebox{\linewidth}{!}
{   \setlength\tabcolsep{0.5pt} 
    \begin{tabular}{|c|c|c|c|c|c|c|c|c|c|c||c|c|}
    \hline
    \textbf{Devices} &  \textbf{Metrics}
    & \textbf{2VXT} & \textbf{3VLB} & \textbf{2A1A} & \textbf{2GTP} & \textbf{2X9A} & \textbf{1RKE} & \textbf{4GAM} & \textbf{4JCV} & \textbf{4LW4} & \textbf{Max} & \textbf{Avg.} \\ \hline\hline
    
    \multirow{2}{*}{CPU} 
    & Speedup & 1$\times$ & 1$\times$ & 1$\times$ & 1$\times$ & 1$\times$ & 1$\times$ & 1$\times$ & 1$\times$ & 1$\times$ & 1$\times$ & 1$\times$ \\ \cline{2-13}
    & Throughput & 72.5 & 55.5 & 86.5 & 94.4 & 432.2 & 100.352 & 15.9 & 28.5 & 42.8 & 432.2 & 103.2 \\ \hline \hline

    \multirow{3}{*}{A100} 
    & Speedup & 9.5$\times$ & 9.5$\times$ & 10.6$\times$ & 10.1$\times$ & 9.9$\times$ & 9.9$\times$ & 8.9$\times$ & 9.3$\times$ & 9.2$\times$ & 10.6$\times$ & 9.7$\times$ \\ \cline{2-13}
    & Throughput & 691.9 & 525.6 & 916.6 & 954.5 & 4281.0 & 989.8 & 142.0 & 266.0 & 392.1 & 4281.0 & 1017.7 \\ \cline{2-13}
    & Chunk Division & \usym{2717} & \usym{2717} & \usym{2717} & \usym{2717} & \usym{2717} & \usym{2717} & \ding{52} & \ding{52} & \usym{2717} & - & - \\ \hline \hline
    
    \multirow{3}{*}{H100} 
    & Speedup & 17.5$\times$ & 18.6$\times$ & 19.2$\times$ & 19.9$\times$ & 20.0$\times$ & 19.4$\times$ & 18.2$\times$ & 18.5$\times$ & 18.7$\times$ & 20.0$\times$ & 18.9$\times$ \\ \cline{2-13}
    & Throughput & 1269.2 & 1031.0 & 1660.2 & 1875.2 & 8652.4 & 1946.6 & 289.5 & 527.0 & 802.0 & 8652.4 & 2005.9 \\ \cline{2-13}
    & Chunk Division & \usym{2717} & \usym{2717} & \usym{2717} & \usym{2717} & \usym{2717} & \usym{2717} & \ding{52} & \usym{2717} & \usym{2717} & - & - \\ \hline
    
    \end{tabular} 
}

\end{minipage}
\hfill
\begin{minipage}{0.34\linewidth}
\centering
\includegraphics[width=\linewidth]{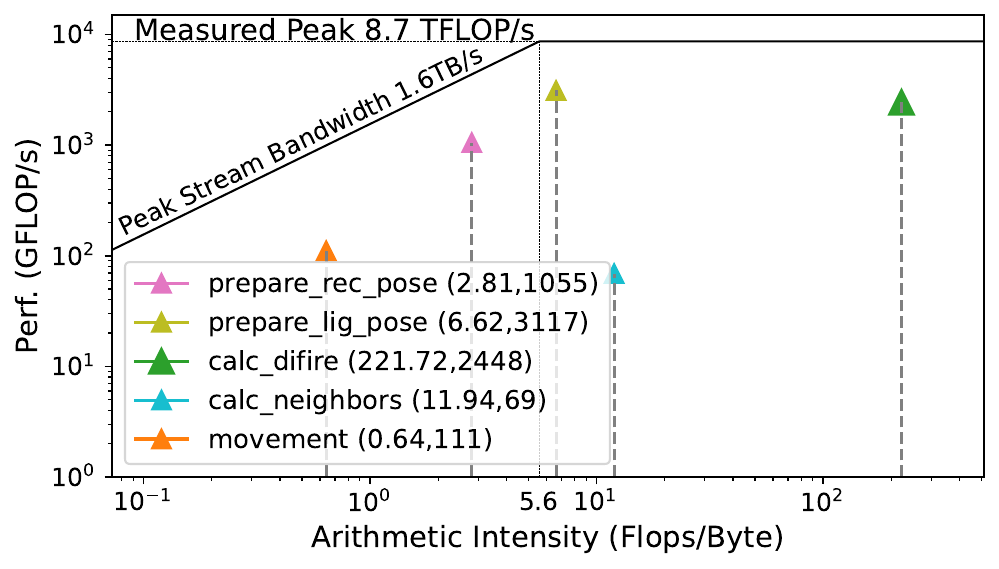}
\caption{Roofline analysis of \method{} kernels at double precision on A100.} \label{fig:roofline}
\end{minipage}
\end{figure*}

\begin{figure*}[!t]
    \centering
    \includegraphics[width=\textwidth]{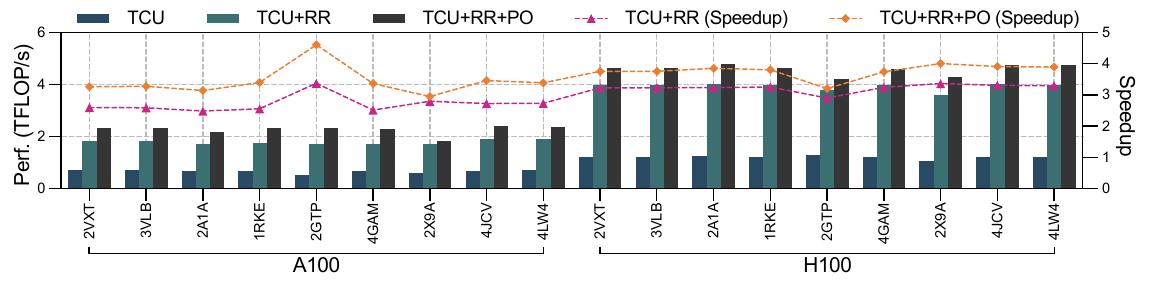}
    \caption{Step-wise breakdown of the impact of different optimizations applied to the hotspot \textit{calc\_dfire} kernel. Here, TCU denotes the TCU-based reformulation, RR denotes register remapping, and PO denotes pipeline overlapping.}
    \label{fig:breakdown}
\end{figure*}

\textbf{Baselines.} 
We compare \method{} with LightDock-Rust, an optimized CPU implementation of LightDock that is widely used in applications such as drug discovery~\cite{chatzifrangkeskou2025atr}, structural prediction~\cite{eronen2024structural}, and membrane protein modeling~\cite{roel2020integrative}. LightDock-Rust achieves an 8--10$\times$ speedup over the original Python implementation~\cite{roel2020integrative} and serves as a strong CPU baseline. 
We further devise a CUDA core implementation of the energy score computation to generalize the \method{} framework to GPUs without FP64 TCU support. 
This implementation preserves the same computational formulation but is executed entirely on standard CUDA cores, also serving as a controlled GPU baseline to evaluate the performance of the TCU implementation. 
All methods are evaluated using FP64 precision to ensure numerical consistency with LightDock and enable a fair comparison across CPU, CUDA-core, and TCU-based implementations.


\textbf{Datasets.} The Protein-Protein Benchmark 5 and Affinity Benchmark Version 2 (BM5.2)~\cite{vreven2015updates} are used to evaluate our framework, which consists of 55 realistic unbound structures:
antibody-antigen (A); enzyme–inhibitor (EI); enzyme–substrate (ES); enzyme complex with a regulatory or accessory chain (ER); others, G-protein containing (OG); others, receptor containing (OR); others, miscellaneous (OX). We select nine representative spanning all category for performance assessment, including docking tasks with smaller computational scales, like 2X9A, and larger docking tasks like 4GAM. Tab.~\ref{tab:dataset} shows the details of these complexes.

\textbf{Environments.} For baseline LightDock-Rust, we test on two 10-core Intel Xeon Silver 4210 CPUs with 40 threads. For \method{}, we test an NVIDIA H100-80GB GPU workstation and a multi-GPU system of up to 800 NVIDIA A100-40GB GPUs (200 nodes, each with 4 GPUs). 
We use CUDA 12.4 to compile \method{}.

\subsection{End-to-End Performance Evaluation on a Single GPU }

In this section, we evaluate the end-to-end docking performance of \method{} on nine representative structures using NVIDIA A100 and H100 GPUs. Since the LightDock-Rust baseline only supports CPU execution, we run it with 40 CPU threads. For all methods and structures, each docking simulation comprises 100 steps, and we report the mean end-to-end execution time across 5 times repeated runs(after 2 times warm-up).
Tab.~\ref{tab:summary} shows the detailed speedup and the computation throughput of \method{} compared to the baseline. According to the results, 
\method{} achieves an average speedup of 9.7$\times$ and 18.9$\times$, with corresponding throughputs of 1017.7 and 2005.9 agents/s on A100 and H100 GPUs, respectively, compared to LightDock-Rust.
Meanwhile, for larger protein docking tasks like 4GAM, 4JCV, the automatic out-of-core chunk division approach can effectively split the workload into smaller chunks, preventing OOM errors during docking.

\begin{figure}[!t]
    \centering
    \includegraphics[width=\columnwidth]{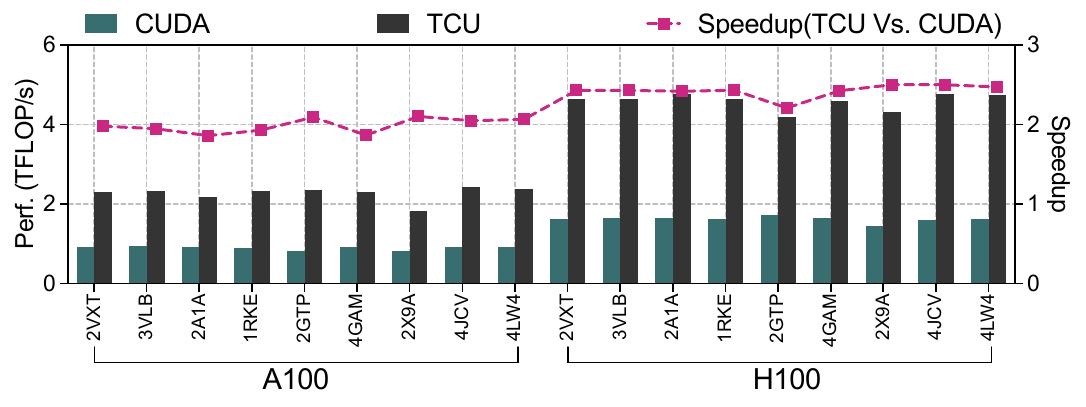}
    \caption{Performance improvement of TCU-accelerated \textit{calc\_dfire}(TCU) compared to CUDA core implementations(CUDA).}
    \label{fig:breakdowntc}
\end{figure}

\begin{figure*}[!t]
    \centering
    \begin{minipage}[t]{0.70\textwidth}
        \centering
        \includegraphics[width=\textwidth]{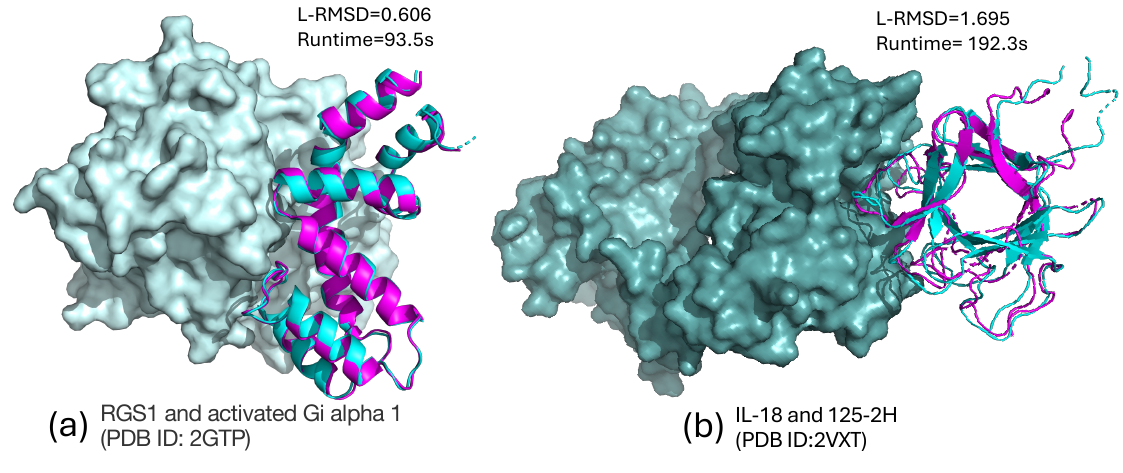}
        \caption{\method{}'s prediction, run on A100 GPU. The magenta ligand is experimental (X-ray crystallography); cyan is the prediction.
        }
        \label{fig:image1}
    \end{minipage}
    \hfill
    \begin{minipage}[t]{0.27\textwidth}
        \centering
         \includegraphics[width=\textwidth]{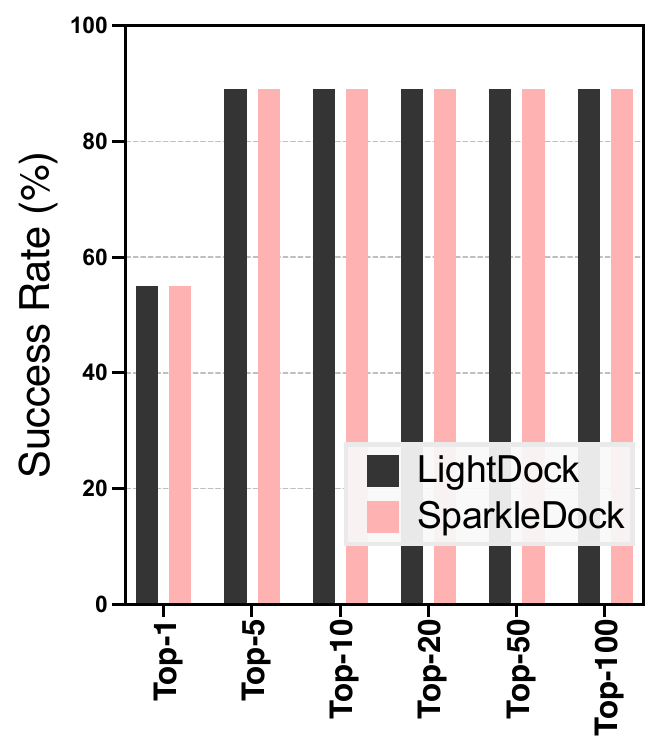}
        
    \caption{Success rate comparison versus LightDock~\cite{roel2020lightdock}.}
    \label{fig:image2}
    \end{minipage}
\end{figure*}

\subsection{Roofline Analysis of CUDA kernels}

We further measure the peak performance of the designed CUDA kernels. 
We use the representative \textit{2VXT} to perform one-step simulation and carry out roofline analysis generated by NVIDIA Nsight Compute (NCU) for these CUDA kernels on an A100 GPU. According to Fig.~\ref{fig:roofline}, \textit{prepare\_lig\_pose} and \textit{calc\_dfire} are the compute-bound kernels achieving over 25\% of the measured peak performance. The \textit{prepare\_rec\_pose} is the memory-bound kernel, achieving 92\% memory occupancy according to the NCU. The performance of remaining kernels \textit{movement} and \textit{calc\_neighbors} is lower than others. The reason is that the total computations of these two modules are quite small, and the launched thread blocks are insufficient to saturate the available resources on A100.
However, these kernels are not time-intensive, consuming less than 1\% of the time, which results in a negligible impact on overall performance.




\subsection{Breakdown Analysis of Energy Score Calculation}
For the bottleneck \textit{calc\_dfire}, we implement a fused TCU kernel design (\texttt{TCU}), register remapping for TCU(\texttt{RR}), and pipeline overlapping (\texttt{PO}). To evaluate the performance improvement of each approach, we conduct a step-by-step breakdown analysis on A100 and H100 GPUs to compare the performance improvements of each design for nine datasets. Fig.~\ref{fig:breakdown} shows the detailed evaluation results.

\textbf{Effectiveness of fused TCU kernel design.}
After applying the \texttt{TCU} design, the \textit{calc\_dfire} kernel achieves 669 (A100) and 1,220 (H100) GFLOP/s peak performance. Compared with the theoretical peak performance of A100 and H100, these results still show a performance gap. Further analysis revealed that the main cause of this gap to be the misalignment between the tensor core and CUDA core patterns at the register level, which introduces additional shared memory movement overhead.


\textbf{Effectiveness of register remapping for TCU.} By further incorporating \texttt{RR}, \textit{calc\_dfire} achieves up to approximately 2.7$\times$ and 3.2$\times$ speedups over the \texttt{TCU} design on the A100 and H100 GPUs, respectively, reaching computational throughputs of 1.80 TFLOP/s on the A100 and 3.93 TFLOP/s on the H100. These results indicate that using \texttt{RR} can effectively align the sum of squares computing pattern with the FP64 TCUs at the register level. By keeping intermediate operands and results in registers, \texttt{RR} avoids redundant data transfers through shared memory, reduces data-movement overhead, and more fully exploits the computational capability of FP64 Tensor Cores.


\textbf{Effectiveness of pipeline optimization.} By further applying \texttt{PO}, \textit{calc\_dfire} achieves up to approximately 1.3$\times$ and 1.2$\times$ speedup compared to the \texttt{RR} implementation on A100 and H100 GPUs, achieving the computational throughputs of 2.27 TFLOP/s and 4.60 TFLOP/s, respectively. The above results demonstrate that \texttt{PO} further improves the kernel performance by overlapping data transfers from global memory to shared memory with computation.

\subsection{Performance Evaluation of TCU Reformulation}
In this section, we further evaluate the performance of the proposed TCU-accelerated \textit{calc\_dfire} kernel. Fig.~\ref{fig:breakdowntc} shows the detailed performance comparison between the TCU implementation and the CUDA core implementation. According to the results, the TCU-accelerated kernel achieves around 1.99$\times$ speedup on single A100 GPU and 2.43$\times$ acceleration on single H100 GPU. These results demonstrate that the proposed TCU reformulation can significantly enhance the bottleneck \textit{calc\_dfire} kernel. The performance gain on the H100 GPU is more pronounced. We analyze that the Hopper SM architecture executes Tensor Core MMA instructions and
auxiliary CUDA Core instructions more efficiently, potentially improving their overlap and reducing pipeline stalls.


\subsection{Docking Accuracy}

We evaluate the docking accuracy of \method{}. We first conduct the docking visualization evaluations for the 2GTP and 2VXT datasets as shown in Fig.~\ref{fig:image1}. The magenta ligands represent the experimentally determined structures obtained by X-ray crystallography, whereas the cyan ligands represent the conformations predicted by \method{}. For both complexes, the predicted ligand conformations closely align with the corresponding experimental structures, indicating that \method{} can accurately reproduce the native binding poses. 
We quantify the docking accuracy using ligand root-mean-square deviation (L-RMSD), which measures the structural deviation between the predicted ligand pose and the corresponding ground-truth pose determined by X-ray crystallography. The L-RMSD values for both 2GTP and 2VXT are below 2.0~$\mathring{A}$, demonstrating that the predicted conformations are highly consistent with the experimentally determined structures. 

In addition, we evaluate the docking success rates of the nine selected complexes at different ranking thresholds using DockQ~\cite{mirabello2024dockq}. According to Fig.~\ref{fig:image2}, the success rates at top ranks 1, 5, 10, 20, 50, 100 of \method{} reach 55.6\%, 88.9\%, 88.9\%, 88.9\%, 88.9\%, and 88.9\%, respectively. These results are identical to those obtained by the LightDock implementation. Overall, the visualization, L-RMSD, and DockQ results demonstrate that \method{} preserves the docking accuracy of LightDock while substantially improving its computational performance.




\subsection{Validation of the Performance Model}
To evaluate the accuracy of $T_{total}$, we use the representative 2VXT, 4GAM, 4LW4, and 4JCV complexes to conduct the experiments on 16 A100 GPUs. According to Tab.~\ref{tab:perfaccuracy}, the predicted docking time closely matches the measured time, with the mean absolute percentage errors (MAPEs) of 12.55\%. Despite minor prediction errors caused by runtime factors such as MPI initialization and resource setup overheads that are not explicitly captured by the model, $T_{\mathrm{total}}$ provides sufficiently accurate estimates to guide the workload distribution strategy described in Sec.~\ref{sec:load}.

\begin{table}[]
\renewcommand{\arraystretch}{1.2}
\caption{Model accuracy validation.}
\resizebox{\linewidth}{!}
{
\setlength\tabcolsep{1.1pt} 
\begin{tabular}{|c|c|c|c|c|c||c|}
\hline
Dataset   & $T_{dfire}$(s)  &   $T_{pose}$(s)        & $T_{other}$(s)      &$T_{data}$(s) & $T_{total}$(s)   & Measured runtime(s) \\ \hline \hline
2VXT  & 13.70 & 0.39 & 0.02  & 0.03 &   14.13   & 17.107  \\\hline
4GAM  & 215.99 & 5.94 & 0.05 & 0.39 &  222.37    & 198.934  \\\hline
4JCV  & 65.97 & 1.29 & 0.03 & 0.09 & 67.37   & 58.6159  \\\hline
4LW4 & 34.28 & 0.99  & 0.02 & 0.07 & 35.35 & 33.318 \\\hline
\end{tabular} \label{tab:perfaccuracy}
}
\end{table}

\begin{figure}[t]
    \centering
    \includegraphics[width=\linewidth]{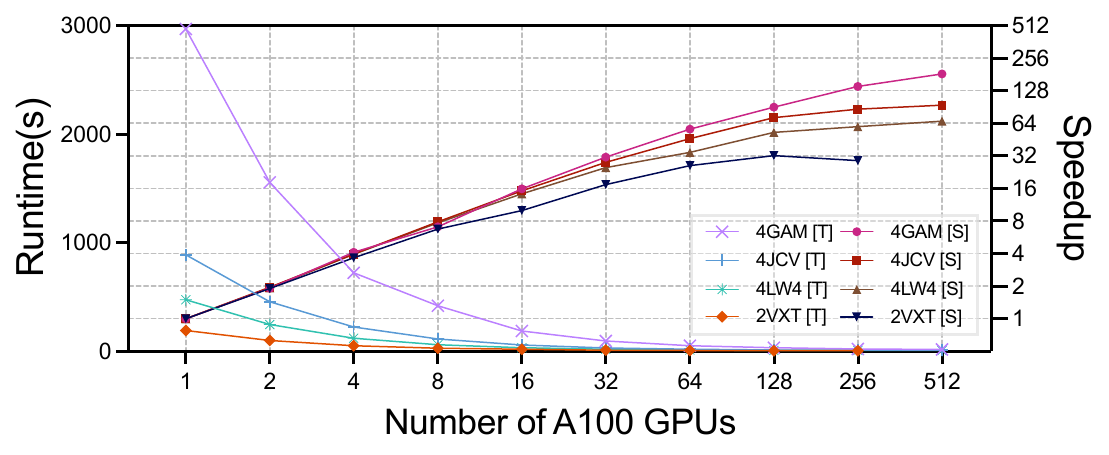}
    \caption{Strong scaling evaluation of SparkleDock on four complexes, where \texttt{T} and \texttt{S} denote the runtime and speedup.}
    \label{fig:scaling}
\end{figure}

\begin{figure}[t]
    \centering
    \includegraphics[width=\linewidth]{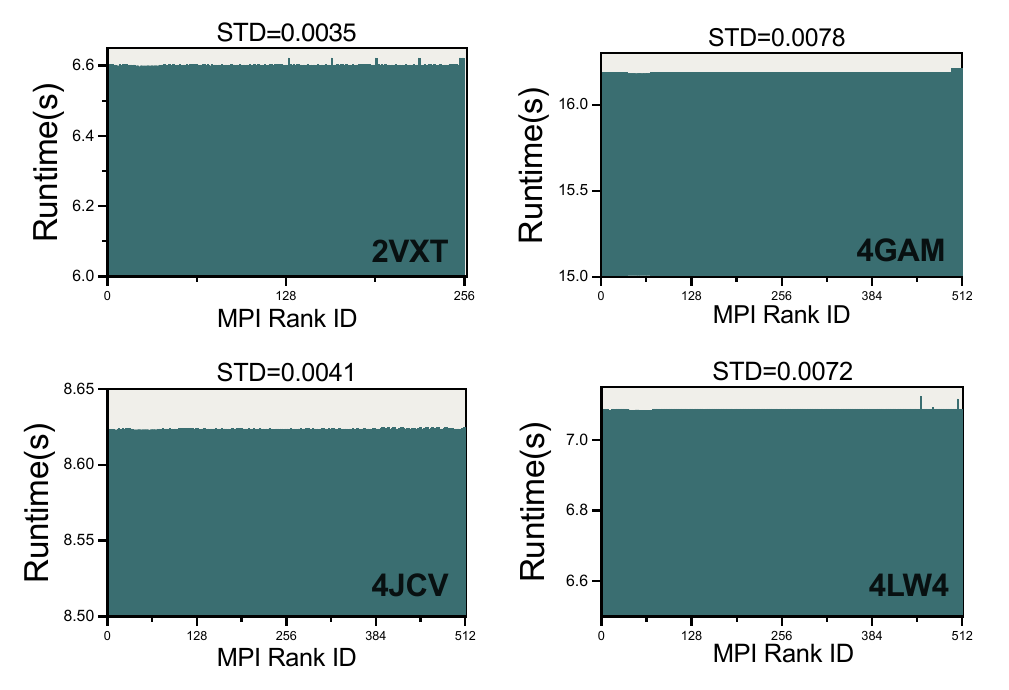}
    \caption{Load balance evaluation of MPI ranks (The green bars denote the runtime in each MPI rank).}
    \label{fig:load_balance}
\end{figure}

\subsection{Scalability Evaluations}

In this section,
we use the 2VXT, 4GAM, 4JCV, and 4LW4 to evaluate the scalability of \method{}.
Fig.~\ref{fig:scaling} shows the strong scaling evaluations for these four complexes.
SparkleDock achieves 183.1$\times$, 94.4$\times$, and 67.0$\times$ speedups for 4GAM, 4JCV, and 4LW4 for 512 GPU scaling, respectively, and a 32.1$\times$ speedup for 2VXT for 256 GPU scaling, relative to a single A100 GPU.
Compared to LightDock-Rust, \method{} delivers over two orders of magnitude speedup, allowing the most flexible macromolecular docking tasks to complete within one minute. For large macromolecules like 4GAM and 4JCV, \method{} adaptively scales across 1 to 4 GPUs without OOM issues, demonstrating the effectiveness of automatic chunk division.
The variation in scaling efficiency among the four complexes is primarily attributed to the differences in workload size. In particular, the 4GAM contains substantially more computation workload, allowing each MPI rank to receive sufficient computation and better overlap initialization, data movement, and synchronization with GPU-accelerated computation. In contrast, the 2VXT has a much smaller workload. In large-scale scaling, the amount of computation assigned to each MPI rank becomes too limited to exploit the GPU acceleration and overlap the fixed communication and synchronization overheads, limiting the scaling efficiency.


We further evaluate the load balance across MPI ranks for the four datasets in Fig.~\ref{fig:load_balance} at the largest evaluated scale in Fig.~\ref{fig:scaling}. We find the computation time of each MPI rank remains consistent, with the standard deviation (STD) of 0.0035, 0.0078, 0.0041, and 0.0072 for the 2VXT, 4GAM, 4JCV, and 4LW4 datasets. These small variations demonstrate that the proposed performance model guided MPI load balancing strategy effectively spreads the workload across MPI ranks, thereby reducing synchronization delays at large GPU scales.

\section{Conclusion}\label{sec:conclusion}

We introduce \method{}, a scalable framework for flexible macromolecular docking. It incorporates efficient design strategies, including fine-grained parallelism, Tensor Core (TCU) mapping, pipeline optimization, and MPI-based scaling.
To our knowledge, \method{} is the first flexible docking framework capable
of multi-GPU scaling while utilizing TCUs, enabling large-scale virtual screening.
Beyond docking, we introduce an efficient design of pairwise distance computation, which is crucial in fields like machine learning and data mining. Future work will extend \method{} to other distributed systems(ARM-based platforms like Fugaku), and explore TCU reformulation to other domains.


\section*{Acknowledgment}
This work was supported by the National Key Research and Development
Program of China (2025YFB4507000), Taishan Scholarship (tstp20240506, tsqn202408087), National Research Foundation Singapore (NRF) and the Ministry of Digital Development and Information (MDDI) under the AI Visiting Professorship (AIVP-2025-005), National Research Foundation Singapore through the National Quantum Office, hosted in A*STAR, under its Quantum Engineering Programme 3.0 Funding Initiative (W24Q3D0002), and Hybrid Quantum-Classical Computing (HQCC) 1.0 Funding Initiative (S24Q7D7001).




\bibliography{reference}
\end{document}